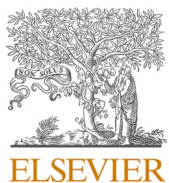
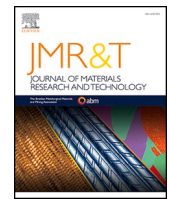

# Microstructure and phase stability within the AlMoNbTiZr system: design tools and compositional boundaries for a high-entropy alloy

M. Casas-Luna [*], D. Preisler, J. Kozlík, J. Stráský

*Charles University, Faculty of Mathematics and Physics, Dept. of Physics of Materials, Ke Karlovu 5, 121 16, Prague, Czechia*

A B S T R A C T

This study explores Ti-containing complex concentrated alloys (CCAs) within the AlMoNbTiZr system, focusing on compositions located in regions of the Bo–Md diagram characterized by low bond order (Bo) and d-orbital energy level (Md). Four alloys were designed near the line predicting stress-induced martensite formation in conventional Ti alloys, then cast, annealed at 1200 °C, and water quenched. Their microstructures and phases were analyzed and compared against phase prediction tools commonly applied to high-entropy alloys (HEAs), namely empirical parameters and CALPHAD simulations. Results highlight the strong influence of chemical affinity, particularly the roles of Al, Mo, and Zr concentrations, on solid-solution stability. A maximum Al content of ~10 at.% was identified as the threshold for achieving a "single-phase" microstructure, observed in the 10Al15Mo10Nb35Ti30Zr alloy. This alloy exhibited a bcc/b2 structure with high compressive strength (>1300 MPa), low Young's modulus (~28 GPa), and limited strain (<6 %), but lacked the transformation-induced strengthening mechanisms expected for Ti alloys with comparable Bo–Md values.

## 1. Introduction

Multi-principal element alloys (MPEAs), also known as complex concentrated alloys (CCAs), have gained considerable attention in the last two decades due to their intriguing microstructures and properties that cover a wide range of applications [1,2]. Since the pioneering studies by Yeh et al., and Cantor et al. on the firsts single-phase MPEAs, named high-entropy alloys (HEAs, referencing the high entropy energy that stabilizes a muti-element solid solution) [3,4]. In this point, it is important to stress that the HEA concept referred exclusively to solid solutions in alloy systems made up of 5 or more elements, each of them in a concentration between 5 and 35 at.% [3]. To distinguish HEAs as a subcategory of CCAs is fundamental in the literature, highlighting their single-phase solid solution. CCAs, in contrast, do not exhibit a single-phase region due to a preferential long-range order of some of their constituents allocated in two or multiple phases and therefore are outside the definition of the high-entropy system [5,6].

This is the reason for utilizing numerous physicochemical and thermodynamic analyses to predict the formation of solid solutions and/or intermetallics within the CCA systems. Up to now, the design and phase prediction of CCAs represent an unresolved challenge that has made significant progress with the help of empirical parameters, computed phase diagram calculations (Calphad) and machine learning approaches [7–13].

HEAs made of 3d transition metals typically form face-centered cubic (*fcc*) structures, expanding the austenite phase field in the Fe–Ni–Cr system. This is due to their similar electronegativities, atomic size, and low mixing enthalpies. In contrast, refractory elements, including elements from Group IV, V, and VI, have more varied electronegativities and atomic sizes, arranged typically in body-centered cubic (*bcc*) and hexagonal close-packed (*hcp*) structures. The last group of elements are characterized by their refractory properties (Mo, Ti, V, Nb, Hf, Ta, Cr, and W). Solid solutions of CCAs that contain some of these elements are known as refractory high-entropy alloys (RHEAs), which have demonstrated high strength, oxidation, corrosion, and wear resistance at high temperatures in comparison to traditional alloys [14–21].

Design of Ti rich RHEAs can benefit from solid and vast design knowledge that can be used as a starting tool in the development of CCAs with a significant amount of Ti within the system [14,15]. For example, the molecular orbital method is one of the most widely used tools for the design of traditional Ti alloys. This method uses the bond order (Bo) and the mean d-orbital energy level (Md) values between Ti and the alloying elements for a specific structural configuration. The Md is related to electronegativity and metallic radius of elements, while the Bo is a measure of the covalent-bond strength between titanium and the alloying elements. Once these mean values of any Ti-containing alloy are calculated, the alloys can be positioned within the empirical Bo-Md diagram, which delineates regions for different structural configurations, namely hcp (α-alloys) and bcc (β-alloys), and shedding light on their deformation mechanisms [22,23]. The transition between the β-alloys and the α+β-alloys regions is characterized by martensitic transformations during mechanical deformation, usually marked by a "Ms = RT" line in the diagram. The region proximal to the Ms line has emerged as an empirical tool for designing enhanced Ti-based alloys, as alloys positioned nearby often exhibit twinning or phase transformation during deformation (TWIP or TRIP effects), what enhances plasticity effects [24,25]. Using this approach, Ti-rich RHEAs have been designed following an extrapolation of the Ms = RT line in the Bo-Md diagram, finding a region in the diagram for Ti-containing HEAs with a bcc structural configuration that exhibits similar deformation mechanisms predicted in traditional Ti alloys [26–28]. The Ti–CCAs region exhibiting TRIP/TWIP effects is located at high values of Bo and Md, far from traditional Ti-alloys (Bo values between 2.75 and 2.90 and Md values between 2.2 and 2.45). This is due to the higher presence of refractory elements that increase the Md and Bo values. This opens a quest on the utility of the Bo-Md diagram as a tool to predict mechanical deformation of Ti-containing CCAs, as well as to identify possible regions inside the diagram that would facilitate such predictions.

The primary objective of this study was to explore Ti-containing CCAs using the molecular orbital approach. Alloy compositions were selected solely based on their position in the Bo–Md diagram, targeting regions close to traditional Ti alloys and near the upper proximity of the Ms = RT line. The AlMoNbTiZr system was chosen because the inclusion of aluminum allowed exploration of different regions within the Bo–Md diagram, while also representing an economical option for lightweight RCCAs [16,29]. Following this selection, four AlMoNbTiZr alloys were arc-melted and annealed to investigate the effect of chemical composition on phase stability. The experimental results were then compared with commonly used phase prediction tools, including empirical parameters and the Calphad method, providing insight into solid-solution stability.

## 2. Materials and methods

### 2.1. Alloys selection

Four CCAs compositions based on the AlMoNbTiZr system were developed by surveying the average bond order ($\overline{Bo}$) and the average d-orbital energy level ($\overline{Md}$) values. The selected compositions were located near the martensitic phase transformation region, approaching the martensitic transformation line ($M_s$) at room temperature from the left, with the aim of exploring this typical region of the diagram but this time with Ti-contanining CCAs (Fig. 1). At this stage, no additional design criteria were applied to ensure that the alloys would form a single-phase solid solution; the selection was based solely on their position in the Bo–Md diagram. The compositions ranged from 20 to 35 at.% Ti, varying Al/Zr ratios with respect to the β-stabilizing elements (Mo and Nb). The nominal compositions and the location of the alloys in the Bo-Md diagram are displayed in Fig. 1 and Table S1 (supplementary material). The alloys were labeled according to their Ti content. It should also be noted that the Al content was kept at 20 at.% in three of the alloys, while in the alloy with the highest Ti content (35 at.%), Al was reduced to 10 at.%, to ensure their proximity to the Ms = RT line.

### 2.2. Casting and thermal processing of AlMoNbTiZr alloys

Pure metals in the shape of small chips with purity above 99.8 % were used for the melting of the CCAs. The alloys were prepared by weighing the pure elements according to the target stoichiometries to get 7.5 g of each composition. The metal mixtures were placed in an arc-melting furnace with a water-cooled copper crucible and a tungsten electrode. The melting of the mixtures was performed under Ar atmosphere. A maximum current of 250 A was applied to melt and mix the metals. The procedure was repeated, flipping the buttons a minimum of 5 times, to ensure the homogenization of the alloys.

The as-cast buttons were cut in half for further heat treatment. The CCAs were coiled in Mo wire (0.1-mm thickness) and wrapped in Zr foil to avoid oxygen intake as much as possible. All alloys were homogenized at 1200 °C for 24 h under an Ar atmosphere, followed by water quenching to retain the high-temperature microstructure.

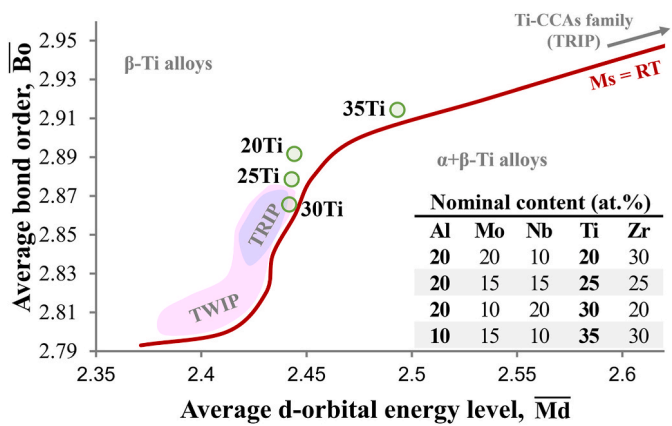

Fig. 1. Alloys selection strategy. Plotting of the AlMoNbTiZr alloys inside the Bo-Md diagram (nomenclature based on Ti content). Table includes the nominal element composition. The empirical phase regions divided by the martensite-start line at room temperature (Ms = RT) are depicted for comparison of these CCAs against traditional Ti-alloys. Similarly, characteristic deformation-mechanism zones and the location of previous Ti–CCAs are pointed out for comparison.

### 2.3. Alloys characterization

*Microstructural and phase composition.* The microstructural observation of the homogenized CCAs was carried out by scanning electron microscopy (SEM) using a FEI Quanta 200F microscope equipped with an energy-dispersive X-ray spectroscopy detector (EDS). Micrographs and elemental analyses were acquired using an accelerating voltage of 20 kV.

The phase identification of the homogenized alloys was performed by Rietveld analysis of the X-ray diffractograms acquired with a Rigaku Rapid II diffractometer (Rigaku, Japan). The X-ray reflections from a Mo-Kα source were collected in a transmission mode by a curved 2D detector and a graphite monochromator using samples with ~250 μm in thickness and a spot size of 0.3 mm. The samples were held on a 3-axes rotating stage, allowing the sample to swing between −15° and 15° in all directions during the measurement to avoid limitations due to very large grains. The diffraction patterns were acquired in a 2θ-angle range of 8–65° with a step size of 0.02° by integrating the 2D rings. The phase identification and pattern fitting were performed using the HighScore Plus software (Panalytical, Netherlands) by comparing the diffraction peaks to suitable ICSD reference patterns.

Additionally, transmission electron microscopy was used to verify the single-phase structure of the 35Ti alloy. TEM analyses were performed using a JEOL JEM-2200FS (Peabody, MA, USA) operated at 200 kV and equipped with an EDAX EDX detector. Disks approximately 0.3 mm thick and 2.5 mm in diameter were cut from the alloy pellet and ground on SiC papers to a final thickness of about 100 μm. The specimens were then electropolished using a Struers Tenupol 5 at 15 V and −30 °C in a 17 % HClO$_4$-methanol solution to prepare TEM foils.

*Microhardness.* The Vickers hardness of the alloys was measured using a microindenter (Qness, Austria). At least 40 indents were made under 1 kgf load applied for 10 s on polished surfaces. Average hardness values are reported with standard deviations. Indentation marks were also examined to correlate the deformation mechanism with the microstructure of the CCAs. In addition, microhardness was related to the fraction content of intermetallic phases, determined by pixel analysis using imageJ 1.53k software (NIH, USA) on at least six SEM-BSE images acquired at different magnifications and locations on the polished surface of each alloy.

*Compression test.* The produced alloys were generally too brittle to be machined into compression test specimens. Only the 35Ti alloy could be processed into square prisms measuring 3.5 mm by side and 5 mm in length, cut using a diamond wafering blade. These specimens were tested by uniaxial compression at room temperature with a digitally controlled testing machine (INSTRON 1186R, Norwood, MA, USA). Due to the material's brittleness, only two specimens produced reliable stress-strain curves, while the other attempts fractured during the initial loading stage. The applied strain rate was $1 \times 10^{-3}\ s^{-1}$. The ultimate compressive strength and the Young's modulus were obtained from the stress–strain curves.

### 2.4. Comparison of microstructure to theoretical phase prediction methods

The physicochemical properties of the individual constituents in any alloy system can provide a fast and easy approach to foresee physicochemical properties and predict microstructure composition in the final alloy. Therefore, in order to evaluate the accuracy of these phase prediction methods, the microstructure of the processed CCAs was compared to the most widely used phase-prediction methodologies.

#### 2.4.1. Empirical parameters

The phase prediction of each mixture was assisted by calculating a list of empirical parameters that have been related to the formation of single-phase solid solutions as a first approach in the development of CCAs. These empirical parameters are usually divided into two categories: i) parameters based on an extension of the Humme Rothery rules (or physicochemical parameters), and ii) parameters based on the system thermodynamics. Among the physical parameters include the relative atomic size difference ($\delta r$), relative Allen's electronegativity difference ($\Delta\chi^A$), valence electron concentration (VEC), and the itinerant electrons per atom ($e/a$). On the other hand, the thermodynamics-based parameters compile those developed to compare the relative influence of enthalpy and entropy in the alloy system. The list of parameters, their classification, and mathematical expressions are summarized in Table S2 of the supplementary material.

#### 2.4.2. Calculation of equilibrium phase diagrams – CALPHAD method

The CALPHAD method was performed in Thermo-Calc software using the "TCHEA4" database, which contains all binary combinations in the AlMoNbTiZr system, as well as some of the ternaries. Such phase diagrams function as a starting point for extrapolations to the complete system. The predicted phase diagrams were constructed between 500 and 2000 °C. The prediction of the phases was monitored at the quenching temperature of 1200 °C and 1300 °C, at which the experimentally observed microstructure has been compared to the predicted equilibrium for each alloy composition.

## 3. Results

The casting of the AlMoNbTiZr system was successfully performed by arc-melting. The following section presents their physicochemical characterization at the heat-treated state of 1200 °C for 24 h followed by water quench.

### 3.1. Microstructure and phase composition

The CCAs containing 20 at.% of Al exhibited a multi-phase microstructure (Fig. 2a–c). The thermodynamic stability of intermetallics arises from the large negative mixing enthalpy between coupled elements ($\Delta H_{mix}$), which weakens the high-entropy effect in the system. Among the constituents, Al is the most reactive, showing the strongest chemical affinity with Zr, followed by Ti and Nb, and the lowest with Mo (Fig. 3a). Contrarily, backscatter SEM images indicate the 10Al15Mo10Nb35Ti30Zr alloy to be a single phase with coarse grains (Fig. 2d). Table 1 summarizes the elemental composition of the phases present in each of the alloys in the quenched state. In general, three phases were qualitatively identified by combining the chemical composition analyses and the X-ray diffraction patterns of the alloys (Fig. 3).

The four CCAs exhibited a solid solution phase containing the five elements arranged in a *bcc* lattice (phase 1 in Figs. 2 and 3b). Most of the Ti and Nb content was present in this solid solution. Additionally, two intermetallic phases rich in Al and Zr were identified in the multi-phase alloys. Both intermetallic phases are not exclusively composed of Al and Zr atoms, as registered by the EDS quantitative analysis (Table 1). The first Al–Zr-rich phase, deficient in Mo content (see Fig. 3c), can be linked to the $Al_3Zr_5$ or $Al_4Zr_5$ intermetallics, both of them having the same structural lattice (hexagonal with a $P6_3/mcm$ space group) and diffraction patterns with some shared peak positions, effect that can contribute to the evident broadening of the diffractograms (phase 2 in Figs. 2 and 3). The second intermetallic (observed only in the 20Ti and 25Ti alloys) is associated to a Laves phase ($AB_2$ structure). The diffraction peaks attributed to the Laves structure appoint to an $Al_2Zr$ intermetallic with atomic substitution of the Al content by either Ti or Mo atoms (phase 3 in Figs. 2 and 3d). The chemical composition of the alloys changes the stoichiometry of their phases, exemplified by the Al/Zr ratio, which seems to be an important parameter to avoid the presence of Al–Zr intermetallics. In general, the $Al_{3–4}Zr_5$ intermetallic (phase 2) maintained an Al/Zr ratio of ~1.0 with very low Mo content (<2.5 at.%). Meanwhile, the Laves phase (phase 3) maintained an Al/Zr ratio of ~0.8, with higher content of Mo and Ti (Table 1 and comparison of Fig. 3c–d).

A deeper quantitative structural description is very complex and restricted due to low diffraction peak intensities of the secondary phases, peaks overlapping, and probable chemical composition variation. This chemical variation is reflected in the displacement of the diffraction peaks, as a result of the lattice-parameters changes, clearly observed in the *bcc* phase of the 35Ti alloy (Fig. 3e).

TEM analysis of the "single-phase" alloy showed a homogeneous elemental distribution and (100) superlattice reflections, confirming ordering within the *bcc* phase. A dark-field image from the (100) reflection revealed finely and uniformly distributed *b2* segregates within the *bcc* matrix (Fig. 4). The TEM-EDS composition was close to the nominal values, with a slightly lower Al content compared to SEM-EDS.

### 3.2. Microhardness and brittle behavior of the alloys

Brittleness of the studied CCAs made it difficult to shape them into mechanical-test specimens. Only the 35Ti alloy was tested under compression and two reliable essays are shown for their mechanical analysis.

The microhardness values of the alloys were in the range of 480–615 HV1 (Fig. 5a). The high content of Al and Mo in the alloys increases the volume fraction of secondary phases up to 73 % in the 20Ti alloy, making the alloy harder and, consequently, more brittle. The impact of intermetallics on the alloys' brittle behavior can be observed in the characteristic indentation marks (SEM-images insets in Fig. 5a). Alloys rich in intermetallics display radial fractures around the indentation marks. In contrast, the "single-phase" alloy shows clean indentation marks without any visible crack propagation, suggesting to some extent, a rather ductile deformation. Compression tests on the 35Ti alloy revealed a high ultimate compressive strength of ~1325 MPa, with a maximum strain of about 6 % before catastrophic fracture. The Young's modulus was estimated at ~28 GPa. Moreover, the compressive stress–strain curves showed no evidence of additional strengthening mechanisms (Fig. 5b).

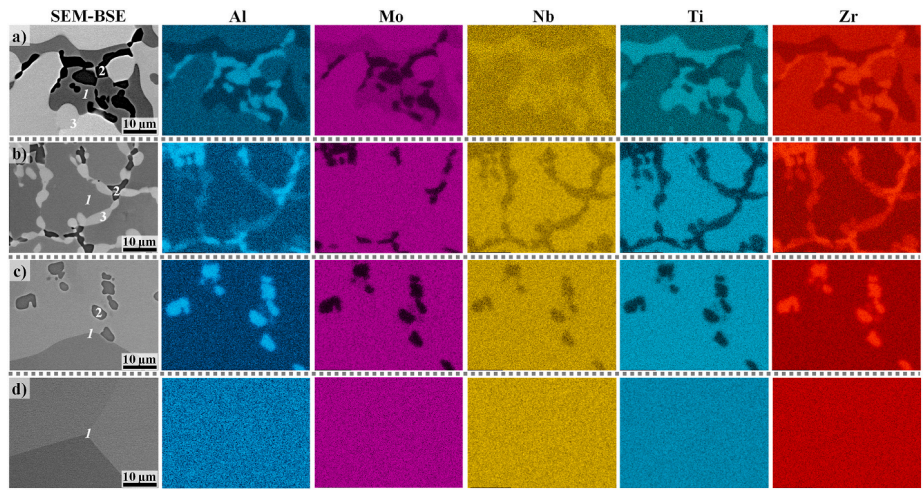

**Fig. 2.** Detail microstructure (SEM-BSE) and element distribution (EDS-mapping) in the CCAs at the quenched state (1200 °C). a) 20Al20Mo10Nb20Ti30Zr, b) 20Al15Mo15Nb25Ti25Zr, c) 20Al10Mo20Nb30Ti20Zr, and d) 10Al15Mo10Nb35Ti30Zr. Three phases can be identified: phase 1, linked to a solid solution of all elements (bcc-phase); phase 2, deficient in Mo; and phase 3, rich in Al–Mo–Zr.

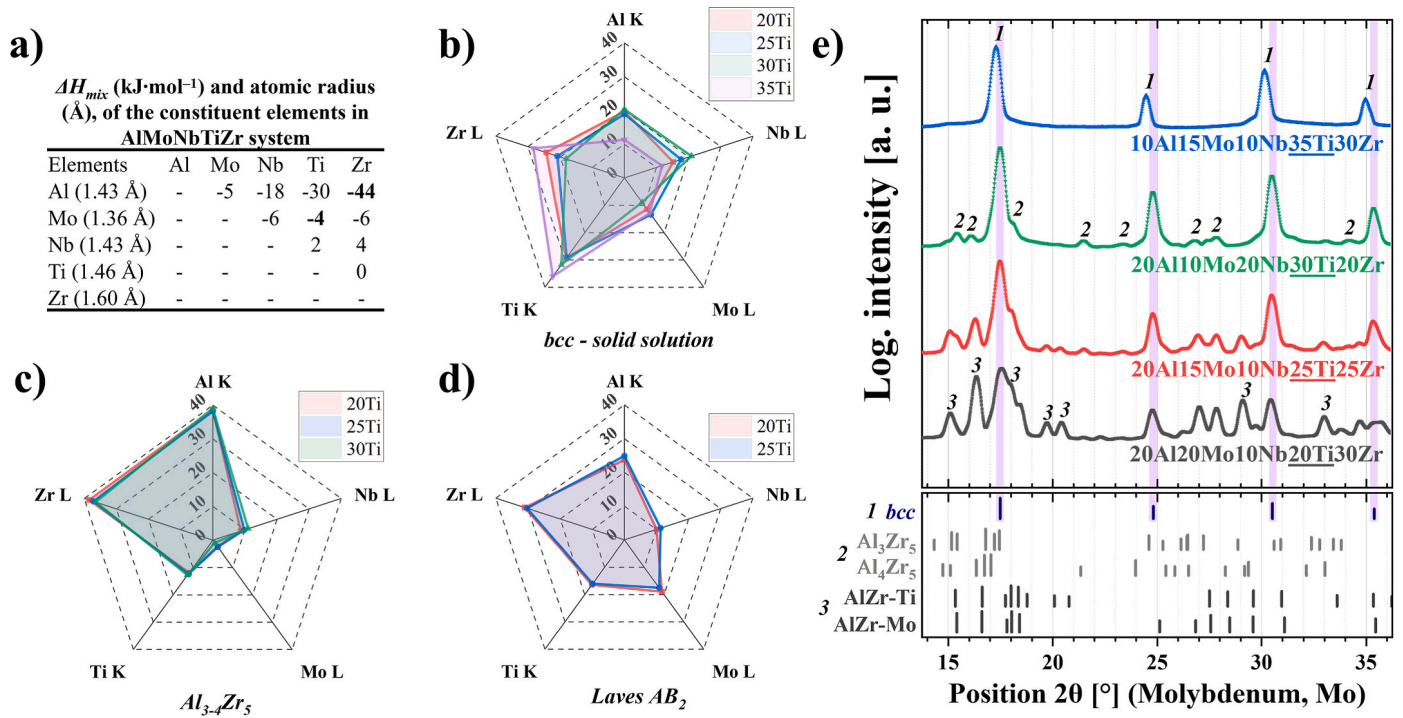

**Fig. 3.** Analysis of phase composition in the AlMoNbTiZr CCAs. a) Mixing enthalpies of the constituent elements and their atomic radii to analyze chemical affinity and coupling effects; b-d) radar graphs showing the composition of each phase present in the alloys; and e) X-ray diffraction patterns and phase identification. Phase 1 (*bcc-solid solution*) (β-Nb; ICDS: 645065), the highlighted bands show the peak positions for this reference phase to illustrate the peak displacement in the HEA; phase 2 (Al–Zr rich phase, $Al_{3-4}Zr_5$) including hexagonal-$Al_3Zr_5$ ($Zr_{3.05}Ti_{1.95}Al_3$, PDF: 04-021-8204), and hexagonal-$Al_4Zr_5$ (ICDS: 167476); and phase 3 (Laves phase with $MgZn_2$ structure) including AlZr–Ti ($ZrTi_{0.3}Al_{1.7}$, PDF: 04-021-8206), and AlZr–Mo (ZrMoAl, PDF: 04-003-9857) compounds. The marks in the lower part of the graph show the reference peak positions of the comparative phases, their lengths are proportional to the theoretical relative intensities.

### 3.3. Calculation of empirical parameters for phase prediction

Based on physical parameters of the elements in the AlMoNbTiZr system, the studied alloys possess low density (ranging between 5.9 and 6.3 g cm$^{-3}$) and melting points above 1375 °C based on Calphad prediction.

1) Hume-Rothery rules: atomic size mismatch, electronegativity, and electronic concentration.

Table 2 summarizes some of the empirical parameters based on an extension of the Hume-Rothery rules applied to the investigated AlMoNbTiZr CCAs. The four CCAs possess an average atomic radii mismatch (δr) below 6.6 %, and a valence electron concentration (VEC) between 4.2 and 4.3. The atomic packing misfitting in the alloys is $\gamma = 1.19$, value beyond the limit to favor a solid solution ($\gamma < 1.175$) [30]. Similarly, the four alloys possess lower $\Delta H^m$ than the stated limit ($-11.6$ kJ mol$^{-1}$) to guarantee a solid solution in CCAs [31,32]. Only the alloy with low Al and Mo content (35Ti) is located in the proximity of this threshold (Fig. 6a). Additionally, the high electronegativity difference calculated in the alloys ($\Delta\chi^A > 9$ %) position them inside the intermetallic formation region (Fig. 6b).

**Table 1**
Chemical composition of the phases present in the microstructure of the alloys at the quenched state. Elemental composition was estimated by punctual energy dispersive X-ray spectroscopy at each phase detected by contrast in BSE-SEM images.

| Nominal alloy composition [at.%] | Phase | Composition [at.%] | | | | | | | Al/Zr |
|---|---|---|---|---|---|---|---|---|---|
| | | Al | Mo | Nb | Ti | Zr | O[a] | N[a] | |
| 20Al20Mo10Nb20Ti30Zr | 1 (bcc) | 19.7 | 11.4 | 15.2 | 29.4 | 24.3 | 0.81 | 0.04 | 0.81 |
| | 2 ($Al_{3-4}Zr_5$) | 38.6 | 2.3 | 8.7 | 11.9 | 38.5 | | | 1.00 |
| | 3 (Laves $AB_2$) | 23.7 | 18.9 | 10.0 | 16.4 | 31.0 | | | 0.76 |
| 20Al15Mo15Nb25Ti25Zr | 1 (bcc) | 18.9 | 13.4 | 17.7 | 29.1 | 20.9 | 0.93 | 0.04 | 0.90 |
| | 2 ($Al_{3-4}Zr_5$) | 38.2 | 2.5 | 9.8 | 12.3 | 37.2 | | | 1.03 |
| | 3 (Laves $AB_2$) | 24.9 | 17.5 | 11.3 | 16.0 | 30.3 | | | 0.82 |
| 20Al10Mo20Nb30Ti20Zr | 1 (bcc) | 20.3 | 8.9 | 20.9 | 31.8 | 18.2 | 1.00 | 0.03 | 1.12 |
| | 2 ($Al_{3-4}Zr_5$) | 39.1 | 1.0 | 11 | 12.7 | 36.5 | | | 1.07 |
| 10Al15Mo10Nb35Ti30Zr | 1 (bcc) | 11.3 | 12.8 | 11.7 | 35.9 | 28.3 | 1.38 | 0.03 | 0.40 |

[a] Oxygen and nitrogen content determined by inert gas fusion.

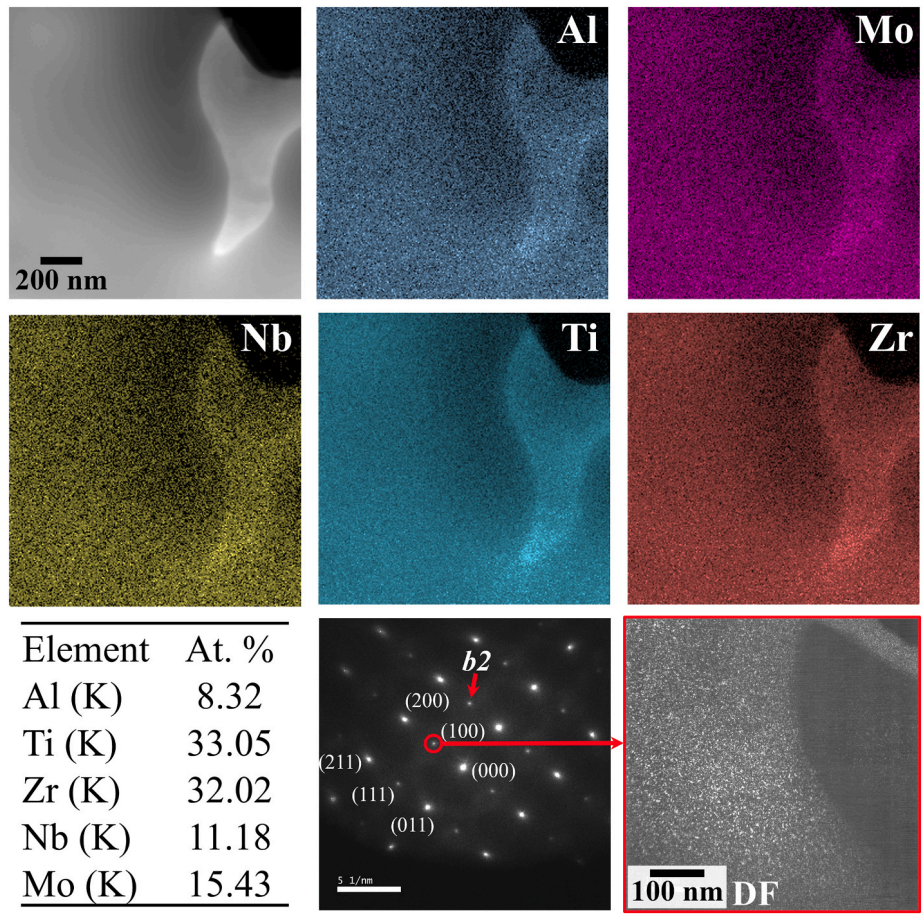

**Fig. 4.** Dark-field TEM image of the 10Al15Mo10Nb35Ti30Zr alloy, showing contrast between thinner and thicker regions of the specimen. Elemental mapping confirms a homogeneous distribution of constituents, with the overall atomic percent composition provided in the inset table. The selected area electron diffraction (SAED) pattern identifies a *bcc/b2* phase, while a dark-field (DF) image taken from the (100) *b2* reflection (red circle in the SAED pattern) reveals finely distributed nanometric *b2* segregates inside the bcc matrix, forming a "single-phase".

2) Thermodynamics-based parameters.

Table 3 compiles most of the thermodynamics-calculation parameters for the four AlMoNbTiZr alloys, while Fig. 6c–f shows a schematic representation of some of these parameters and their phase prediction.

The empirical prediction using the Ω-parameter plotted against δr consistently overestimates the stability of solid solutions; nonetheless, alloys with high Al content (20 at.%) lie near the boundary, suggesting their tendency toward intermetallic formation (Fig. 6c). Differently, other thermodynamic parameters overpredict intermetallic phases, as the experimentally observed "single-phase" alloy (10Al15Mo10Nb35Ti30Zr) is positioned within the intermetallic region (Fig. 6e–f). Despite this, all thermodynamics-based parameters correctly capture the overall trend, placing the single-phase alloy close to the solid solution region and clearly separated from the rest of the studied compositions.

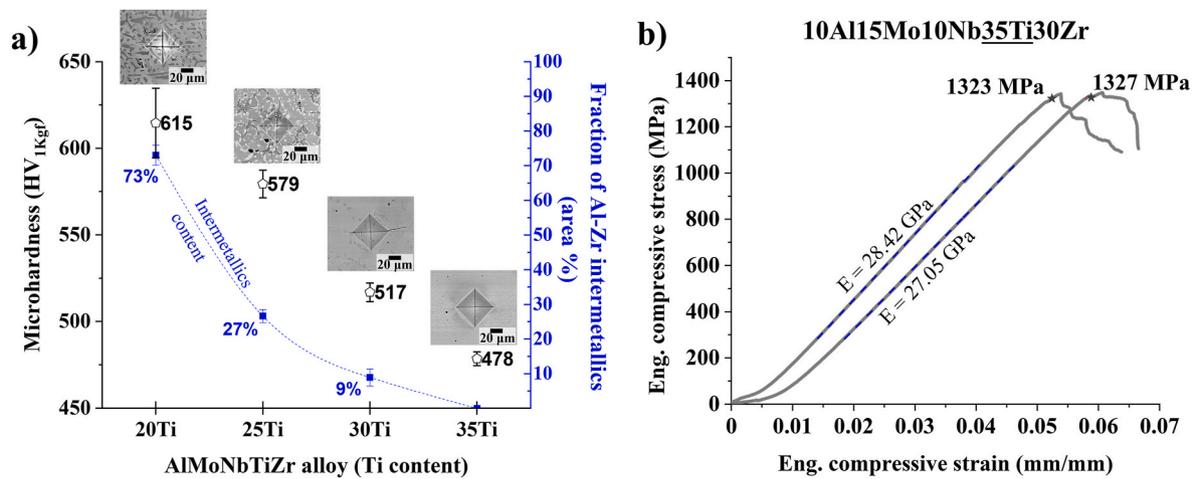

**Fig. 5.** a) Average Vicker's hardness of the alloys in the quenched condition, along with corresponding SEM images of characteristic indentation marks and the average surface fraction of Al–Zr intermetallics. b) Stress-strain curves for the 10Al15Mo10Nb35Ti30Zr alloy, showing the ultimate compressive strength and calculated Young's modulus.

**Table 2**
Physicochemical parameters (based on the Hume-Rothery alloying rules) of the studied alloys.

| System | Density (g/cm3) | δr (%) | γ (−) | Δχ$^A$ (%) | VEC (−) | e/a (−) | T$_m$ (Thermocalc, K) |
| --- | --- | --- | --- | --- | --- | --- | --- |
| 20Al20Mo10Nb20Ti30Zr | 6.3 | 6.07 | 1.19 | 11 | 4.3 | 1.9 | ~1653 |
| 20Al15Mo15Nb25Ti25Zr | 6.11 | 5.56 | 1.19 | 10 | 4.25 | 1.9 | ~1738 |
| 20Al10Mo20Nb30Ti20Zr | 5.93 | 4.97 | 1.19 | 10 | 4.2 | 1.9 | ~1777 |
| 10Al15Mo10Nb35Ti30Zr | 6.19 | 5.71 | 1.19 | 9 | 4.3 | 1.85 | ~1653 |
| **Solid Solution condition** | - | <6.6 | <1.175 | ≥3; ≤6 | - | - | - |

### 3.4. CALPHAD prediction

Fig. 7 plots the CALPHAD method phase prediction for the CCAs. A solid solution in a *bcc* crystal lattice is predicted only in the alloy with 10 at.% of Al between 1275 °C and 1380 °C (predicted melting point of the alloy). For the alloy compositions containing 20 at.% of Al, a multiphase microstructure is expected. At high temperatures, the CCAs comprises a *bcc*-phase alongside an intermetallic phase resembling the $Al_3Zr_5$ compound with a volume fraction around 0.45. Additionally, the calculation predicts formation of a Laves phase with an Al/Zr ratio of 0.67 for the alloy with the lowest Ti content (20 at.%). The volume fraction of the intermetallics in the AlMoNbTiZr system appears to have a correlation to the Al/Zr ratio. By comparison, a single *bcc* phase is only favored at elevated temperatures when the Al/Zr is below 0.33, as exemplified by the 10Al15Mo10Nb35Ti30Zr alloy.

Fig. 8 compiles the CALPHAD phase prediction for the 10Al15Mo10Nb35Ti30Zr alloy showing the effect of each elemental concentration with respect to Ti evaluated at 1200 °C (dashed lines) and 1300 °C (solid lines). The graphs highlight the HEA region, where a single *bcc*-phase is predicted at 1300 °C and the composition restriction of 5–35 at.% of all constituents is met.

The results reveal that Al content should not exceed 10 at.%, requiring adherence to the Al/Zr ≤ 1/3 condition to diminish the formation of Al–Zr intermetallics (Fig. 8a–b). The $Al_3Zr_5$ intermetallic is expected in all compositions below 1200 °C. Also, an increase in Nb at the expense of Ti (above ~20 at.% Nb) results in a dual-phase *bcc/b2* alloy (Fig. 8c). Similarly, Mo stabilizes a Laves phase when its content exceeds the 23 at.% (Fig. 8d).

### 4. Discussion

#### 4.1. Microstructure and phase stability within the AlMoNbTiZr system

The AlMoNbTiZr system presents a promising subject for further research due to its low density, around 6.3 g cm$^{-3}$, making it lighter than other lightweight RCCAs [15,16,34]. Additionally, this system includes more cost-effective elements compared to other RCCAs containing Hf and Ta, making it a more economical option for high-temperature applications. The main drawback is the presence of Al–Zr intermetallics in a wide range of compositions and temperatures, limiting the design of HEAs within this system.

Recently, the application of Ti-alloy design principles to Ti-containing CCAs has shown promise in creating high-performance alloys within the HfNbTaTiZr system [26]. The Bo-Md diagram, a well-established tool for traditional Ti alloys, could offer a practical approach for designing Ti-containing HEAs by guiding composition selection and predicting properties [22,26,27]. However, an important restriction is to ensure the solid-solution formation, which cannot be predicted by the Bo-Md diagram, and different strategies must be followed. In the AlMoNbTiZr system, the studied alloys had Bo and Md values similar to β-Ti alloys (*bcc* phase), unlike previously designed Ti-containing CCAs, which had higher values (Fig. 1) [14,26–28]. The aluminum content lowers the values of electronic parameters, allowing exploration near the Ms = RT line close to common Ti-alloys with TRIP/TWIP deformation mechanisms [22,24]. However, this effect was absent in the studied alloys, as they were brittle due to formation of intermetallics. The presence of such intermetallics is not accounted in the molecular-orbital design method, therefore, ensuring stable single solid-solutions is crucial for further correlations between Ti-containing HEAs and their position within the Bo-Md diagram. Even for the 10Al15Mo10Nb35Ti30Zr alloy, which was confined to a *bcc/b2* structure emulating a solid solution (observed by TEM, Fig. 4), strengthening mechanisms such as phase transformation or twinning were not considered (see Fig. 5b).

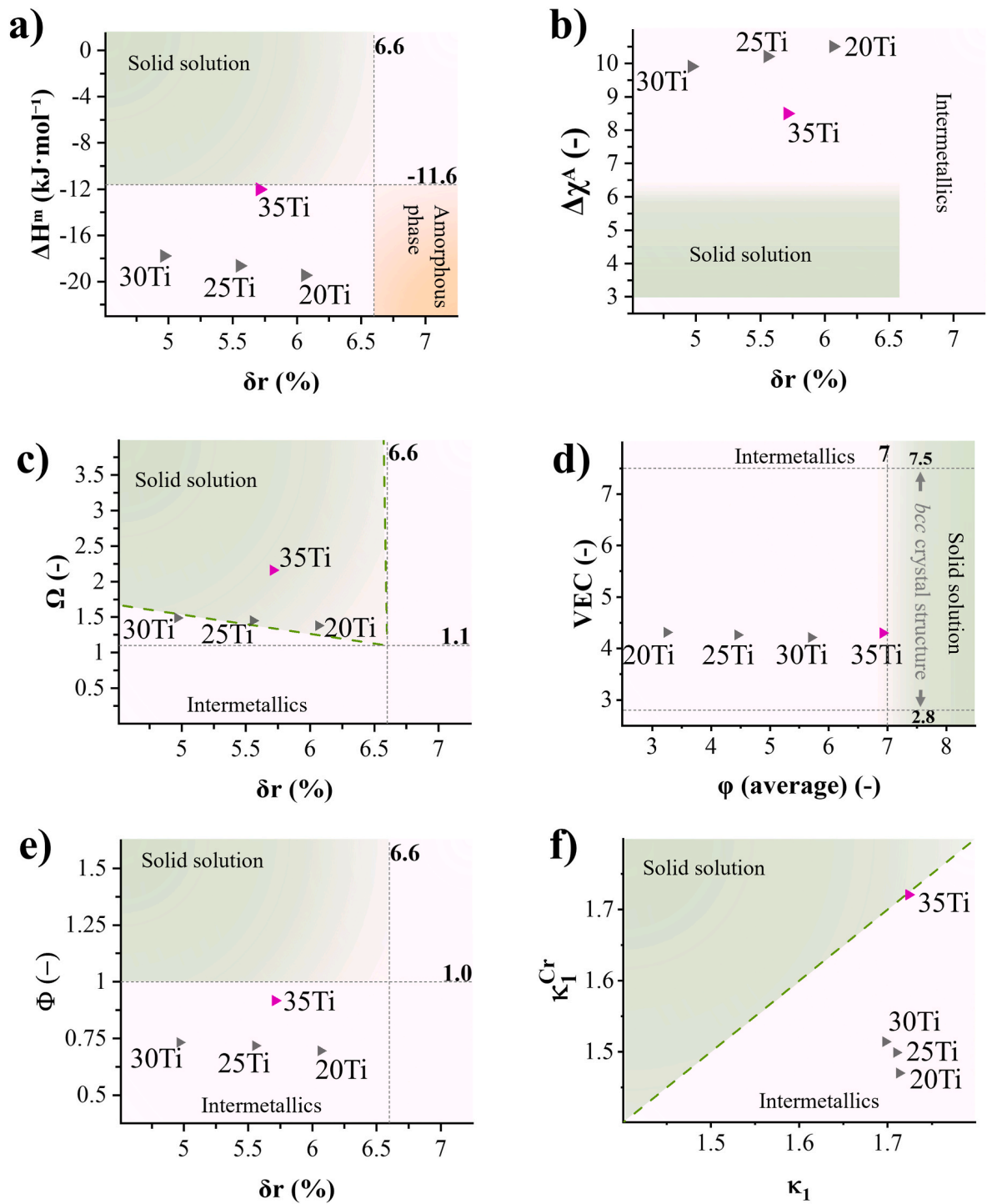

**Fig. 6.** Empirical parameters used for predicting solid-solution stability in CCAs. Parameters based on Hume-Rothery rules: a) Enthalpy of mixing ($\Delta H^m$) and b) electronegativity difference ($\Delta \chi^A$) versus atomic size mismatch ($\delta r$). Parameters based on thermodynamic variables: c) $\Omega$, d) VEC vs $\varphi$, e) $\Phi$, and f) $\kappa_1^{cr}$ vs $\kappa_1$. The green-shaded region highlights the limits and zones for predicting solid solution stability. The AlMoNbTiZr alloys are represented by labeled marks corresponding to their Ti atomic percent concentration.

Similar phase stability to that of the 35Ti alloy has been reported for the $Al_{0.5}NbTa_{0.8}Ti_{1.5}V_{0.2}Zr$ RHEA, where microstructural evolution occurs through spinodal decomposition within the *bcc* phase, couple with *b2* ordering. This highlights the significant changes in phase stability as a function of heat treatment [35].

Despite the "single-phase" *bcc/b2* microstructure in the 35Ti alloy, its mechanical behavior at room temperature showed low ductility but a high ultimate compressive strength (>1300 MPa) and a low Young's modulus (~28 GPa), values typically observed in Ti/Zr/Nb-based alloys that promote a *bcc* phase [26,28,36]. In the other three alloys, the high Al content led to the formation of intermetallics. The strong chemical affinity of Al, measured by its highly negative $\Delta H_{mix}$ with other constituents, induce the segregation of Al in specific thermodynamically stable phases, a behavior commonly observed in Al-containing CCAs [37,38]. This resulted in the formation of mainly two Al–Zr intermetallics (characterized in Fig. 3). These intermetallics significantly increased the hardness of the alloys, when their surface fraction reached 73 %, hardness exceeded 600 HV, compared to 478 HV in their absence (Fig. 5a), representing a much higher specific hardness than other CCAs or conventional Ti alloys used in intermediate-temperature applications [34]. Despite some benefits of the Al–Zr phases, they caused embrittlement in the studied alloys due to their significant phase volume fraction and their spatial arrangement (present at the grain boundaries). Then, it becomes imperative to project the existence and influence of secondary phases on the mechanical properties, to advantageously tailor their distribution and morphology within any CCAs system [39–44].

**Table 3**
Thermodynamic properties and related empirical parameters for predicting solid-solution stability for the AlMoNbTiZr CCAs.

| System | $\Delta H^m$ (kJ·mol$^{-1}$) | $\Delta S^m$ (J·mol$^{-1}$·K$^{-1}$) | $\Delta H^c$ (kJ·mol$^{-1}$) | $\Omega$ (−) | $\Lambda$ (J·mol$^{-1}$·K$^{-1}$) | $\phi/bcc$ (−) | $\Phi$ (−) | $\eta$ (−) | $\kappa_1^{cr} - \kappa_1$ (−) |
|---|---|---|---|---|---|---|---|---|---|
| 20Al20Mo10Nb20Ti30Zr | −19.52 | 12.95 | −33.46 | 1.37 | 0.35 | 4.26 | 0.27 | 0.37 | −0.32 |
| 20Al15Mo15Nb25Ti25Zr | −18.70 | 13.17 | −31.99 | 1.44 | 0.43 | 5.84 | 0.25 | 0.37 | −0.29 |
| 20Al10Mo20Nb30Ti20Zr | −17.84 | 12.95 | −30.30 | 1.48 | 0.53 | 7.47 | 0.24 | 0.37 | −0.27 |
| 10Al15Mo10Nb35Ti30Zr | −12.02 | 12.25 | −20.72 | 2.16 | 0.38 | 9.06 | 0.30 | 0.35 | −0.12 |
| **Solid Solution condition** | ≥ −11.60 | - | - | ≥ 1.1 | ≥ 0.96 | ≥ 7[a] | > 1 | > 1 | > 0 |

[a] Established for light weight alloys [33].

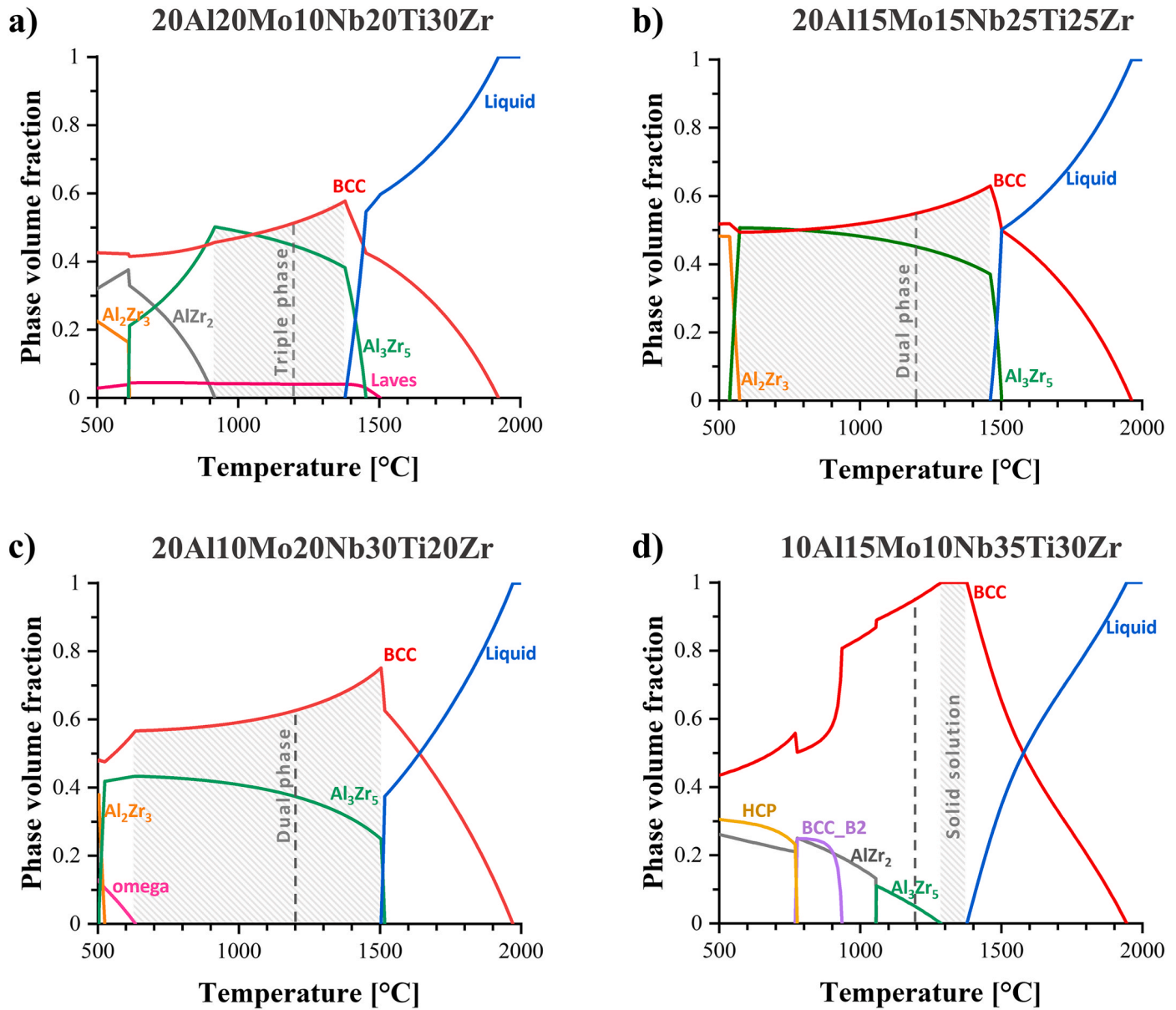

**Fig. 7.** Simulated equilibrium phase diagrams of the four AlMoNbTiZr alloys: a) 20Al20Mo10Nb20Ti30Zr, b) 20Al15Mo15Nb25Ti25Zr, c) 20Al10Mo20Nb30Ti20Zr, and d) 10Al15Mo10Nb35Ti30Zr. The hatched area shows the predicted microstructure composition at high temperature, while the dashed line indicates the quenching temperature.

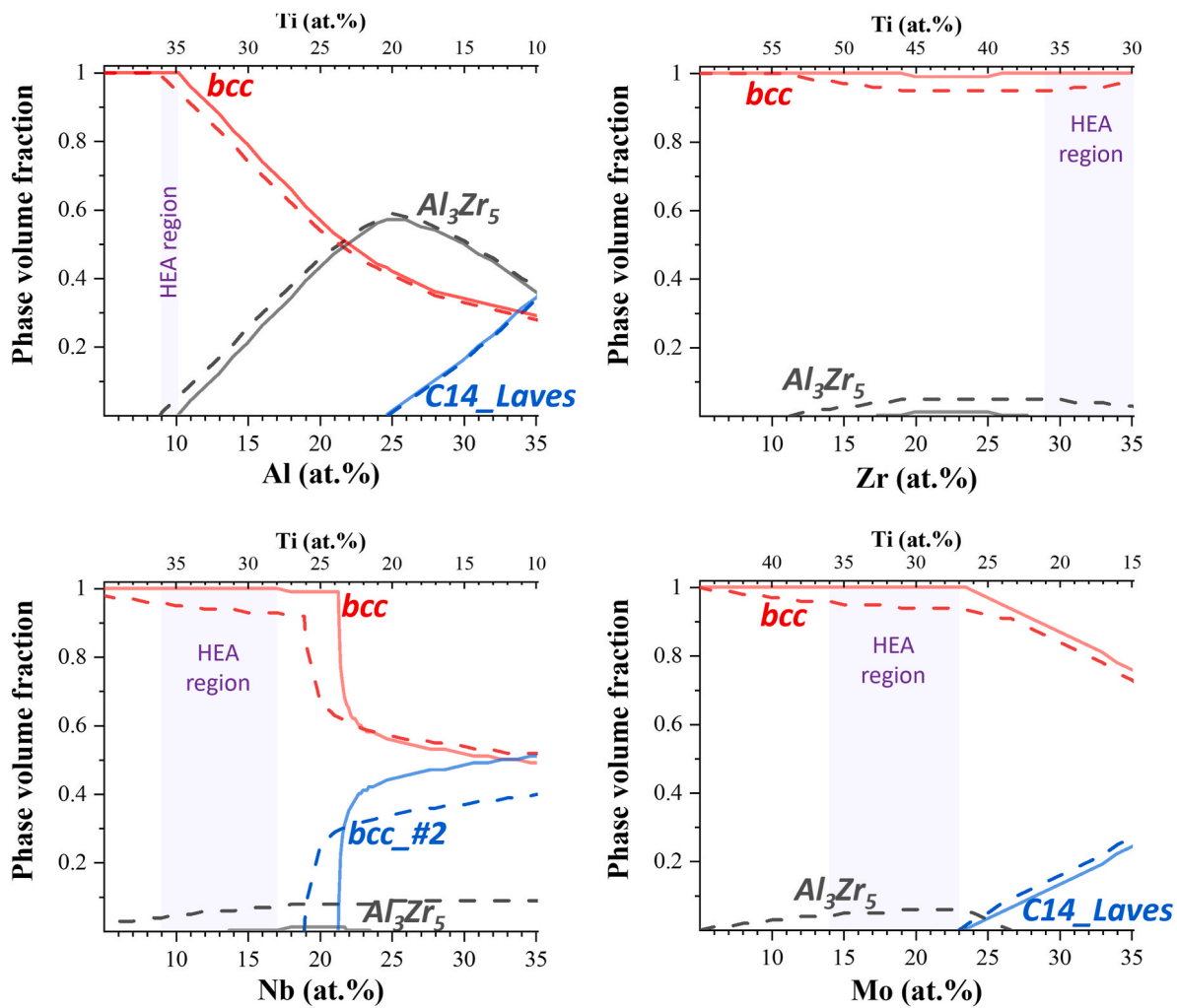

**Fig. 8.** Element composition impact on the phase stability for the 10Al15Mo10Nb35Ti30Zr system estimated by ThermoCalc® (TCHEA4 database). The element composition is adjusted by varying the titanium content in relation to: a) aluminium, b) zirconium, c) niobium, and d) molybdenum, while maintaining the other elements constant in the alloy. Solid lines depict the phase fraction at 1300 °C, while dashed lines are the calculations at 1200 °C. The high entropy zone (HEA region) is identified considering the composition limits that would deliver a single *bcc*-phase at 1300 °C.

### 4.2. Accuracy of Hume-Rothery rules-based parameters on the phase prediction of AlMoNbTiZr system

The simple calculation criteria of empirical parameters constitute their main advantage to be used as a first approach in the CCAs design, compared to other more complex and time-consuming methodologies [8,11,12,45]. However, the parametric approach is generally linked to a low accuracy (generally below 70 %) [8]. Therefore, empirical parameters must be used cautiously and in combination with other alloy-design approaches to increase the accuracy of the phase prediction [10].

In the studied alloys, various correlations involving the lattice distortion (γ-parameter), enthalpy of mixing ($\Delta H^m$), or the electronegativity difference ($\Delta \chi^A$, based on Allen's scale) against the δr anticipate the formation of multiple phases (Fig. 6a-b). These empirical correlations serve exclusively only as starting point for forecasting the stability of solid solutions in various systems [7,8]. Nonetheless, intermetallic phases may persist within regions identified as solid-solution areas by these parameters. This occurs when these phases possess enthalpy of formation close to zero and small atomic size disparities, granting them greater stability compared to a simple solid solution [46]. It is established that the absolute value of $\Delta H_{mix}$ should be close to zero for an element to distribute randomly in a solid solution phase. Positive enthalpy values decrease miscibility, leading to segregation, while more negative values indicate high bonding forces, promoting intermetallics [47].

The electronegativity difference has demonstrated a high precision in predicting elemental segregation. Values of $\Delta \chi_{Allen}$ between 3 and 6 indicate solid solutions, while outside of this range there could be intermetallics [7,48]. Moreover, the $\Delta \chi$ parameter serves as a predictive tool for the formation of Laves phase when $\Delta \chi_{Allen}$ exceeds 7 %, increasing the probability if the atomic size mismatch is greater than 5 % for CCAs with Al content [49]. This applies to the AlMoNbTiZr system, the presence of Laves phases is favored by the high chemical affinity between Al and Zr, as evidenced in microstructural analysis (Fig. 3).

By the side, electronic parameters, *i.e.*, VEC and e/a, are decisive for predicting the crystal lattice structure in alloys. The stability of a given structure depends on the optimal arrangement of electrons, in this way, a crystal structure is favored if it accommodates more electrons in lower energy levels [48,50,51]. The alloys within the AlMoNbTiZr system stabilize a *bcc* solid solution, which is exemplified in the 10Al15Mo10Nb35Ti30Zr alloy (Fig. 2d and 3).

### 4.3. Accuracy of thermodynamics-based parameters on the phase prediction of the AlMoNbTiZr system

The thermodynamic parameters show a similar trend between them, at the same time that correctly appoint the intermetallic formation tendency in the alloy system and locating the Ti35 alloy closer to the solid-solution region.

The only exception was observed for the Ω-parameter, which simplifies the calculations too much by comparing only the entropic term of the solid solution with the enthalpy of formation of the most likely competing phases [47]. Based on this criterion, the four alloys incorrectly qualify as solid solutions (Fig. 6c). However, this result should be cautiously interpreted, given the proximity of the alloys to the intermetallic region and the chemical affinity pairing of the alloy components.

Other thermodynamic parameters predict intermetallic formation, accounting for factors such as lattice distortion (Λ-parameter) [52] or the melting temperature in the enthalpic term, which may outweigh the excess configurational entropy of a solid solution (ϕ-parameter) [1,33,53] (Fig. 6d).

More complete models incorporate the full calculation of free energy of both solid solution and potential intermetallics, resulting in more restricted parameters that consider the minimization of the Gibbs free energy of the system. The Φ parameter, compares the Gibbs free energy of the solid solution with the most probable binary intermetallic or single-element segregation [54] (Fig. 6e). Furthermore, by incorporating into the calculation the annealing temperature, the η-parameter [45,55] and the $\kappa_1^{cr}$ – $\kappa_1$ relation [56] allow the prediction of the temperature at which the solid solution stabilizes. From this point of view, the used quenching temperature (1200 °C) is not enough for stabilization of the solid solution (Fig. 6f).

The enthalpy of mixing between Al and Zr is in absolute value much higher than the entropy in the system: $|\Delta H^{IC}| \gg |T\Delta S_{mix}|$ (see Table 3). However, these parameters, like the previous cases, should represent initial alloy design steps, especially for Al-containing CCAs. For instance, one of the exceptions for the Φ-parameter expresses that aluminum atoms might have preferential ordering within *bcc* or *fcc* crystal structures [57], favoring solid solutions at determined concentrations. This can be the case of the 10Al15Mo10Nb35Ti30Zr alloy, that besides the intermetallic predictions, solid solution was obtained at the quenching temperature of 1200 °C. Similarly, the $\kappa_1^{cr}$ – $\kappa_1$ difference has been calculated assuming a partially ordered intermetallic phase with a sublattice arrangement equal to $(A,B)_1(C,D,E)_3$, having a configurational entropy about 60 % of $\Delta S_{mix}$ for a five-component equiatomic solid solution (thus, $\kappa_2 = 0.6$) [56]. This $\kappa_2$ value applies to a lattice arrangement of $(A,B)_3(C,D,E)_5$, paired with the $Al_3Zr_5$ intermetallic phase existing in the AlMoNbTiZr system at high temperatures. However, this configurational entropy decreases if fewer atoms are involved in the intermetallic phase, as is the case in the studied system, presenting a phase deficient in Mo (phase 2 in Table 1). Thus, the solid solution could be stabilized for different compositions within specific temperature ranges.

Relying solely on parametric approaches has proven insufficient to determine the optimal conditions for stabilizing a solid solution; however, they can be useful as an initial screening in CCAs, especially when combination of several parameters is considered. Here, the 10Al15Mo10Nb35Ti30Zr alloy consistently resides near the solid-solution stability region across most of the parameters, surprisingly showing a correct tendency within the studied system towards the stabilization of a single *bcc* phase.

### 4.4. Accuracy of Calphad phase prediction

An overview of the microstructure and phase analysis suggests that the reduction of the Al content promotes the formation of a single-phase *bcc* alloy within the AlMoNbTiZr system. The solid solution in the 10Al15Mo10Nb35Ti30Zr alloy is expected to exist above 1275 °C (Fig. 7d). Therefore, by quenching the alloy at 1200 °C, approximately 5 vol%. of $Al_3Zr_5$ intermetallic is predicted by the CALPHAD method. However, the stoichiometry of the final alloy slightly deviates from the nominal values, showing a lower Mo and Zr content (see Table 1). These minor compositional variations impact the stability of secondary phases, ultimately stabilizing the solid solution and resulting in a single-phase microstructure. The "single-phase" lies near the boundary for single *bcc/b2* phase stability (see Fig. S1, supplementary material), as predicted by the Calphad method, which is often qualitatively correct. Still, the phase prediction of the alloys by the CALPHAD method presented some divergences from the experimental reality, by omitting the Al–Zr Laves phase within the alloys with 20 at.% of Al. On the other hand, the computation reveals a great deal about the important role of Al content in the AlMoNbTiZr system. The thermodynamic phase modelling restricts the Al concentration to a maximum of 10 at.%, maintaining Al/Zr $\leq 1/3$, to avoid high-temperature Al–Zr intermetallics (Fig. 8a). Comparable Al concentration was determined to yield a solid solution in the FeCoNiCrMnAl system, where above the 8 at.% of Al leads to a dual-phase microstructure [29]. Therefore, it is clear that the high content of Al in refractory CCAs promotes the formation of secondary phases with multiple transition metals, having a high affinity for Zr to stabilize up to ten phases [58,59]. However, such intermetallic phases could become single-phase microstructures at high temperatures in RCCAs [16,35].

A low content of Mo is also essential to avoid the formation of secondary phases in the system. In a similar CCA alloy, the effect of Mo within the TiNbZrTaMo system was found to influence elemental dendritic segregation between two *bcc* phases [60]; this effect could be observed in the studied system, where Mo exhibits preferential segregation in the Ti/Nb-rich phase (with *bcc* structure) and is highly deficient in the $Al_3Zr_5$-like intermetallic (phase 2) in the alloys with high Al and Mo content (Table 1).

Consequently, two facts must be considered to obtain a single *bcc*-phase within the AlMoNbTiZr system: 1) Zr has a high chemical affinity with Al to pair in a wide possibility of intermetallics, the ratio between these two elements must be cautiously selected; and 2) Mo exhibits preferential affinity to Ti, being preferentially located in the *bcc* phase (phase 1 in Table 1). Then, the controlled content of Al, Mo, and Zr is crucial to avoid intermetallic formation in the studied CCA system.

The interplay effect of constituents in a quinary CCA cannot be only assessed by a trial-and-error approach or numerous calculations of empirical parameters. A simpler explored methodology consists of a ternary analysis performed in CALPHAD modelling. The variation of two components with respect to a third one can provide a visible tendency of not only individual element effect but also the interplay of three constituents at once. This approach is illustrated in Fig. 9, where the compositional effect of Al, Mo, and Zr, as the most crucial elements for the formation of intermetallics within the AlMoNbTiZr system, was computed in terms of the *bcc*-phase stability at 1300 °C for an alloy composition of $x$Al-$y$Mo-$z$Zr-10Nb35Ti. This ternary analysis applied to a quinary system offers an easy and straightforward approach for monitoring chemical affinities between the constituents and their collective impact on the solid-solution stability.

With this approach, we could estimate that single *bcc*-phase in the system has a concentration limit for Al of 11 at.% only if the Mo content is also kept low (Fig. 9). However, this low Al and Mo concentrations imply an alloy rich in Zr (given the fixed sum of 55 %), falling outside the high-entropy composition. Then, the optimal concentration of Al would be around 8 at.%; composition at which the Zr and Mo content can be indistinctly interchanged to ensure a solid solution. This last condition suggests that the Al/Zr ratio can exceed the 0.33 value, even at double, *e.g.*, 12 at.% of Zr and 35 at.% of Mo (Fig. 9). However, this assumption would have to be demonstrated, as CALPHAD method sometimes lacks accuracy based on the absence of phase diagram information.

## 5. Conclusion

The predictive capability of the Bo-Md diagram for the mechanical behavior of Ti-containing HEAs with a bcc structure remains uncertain,

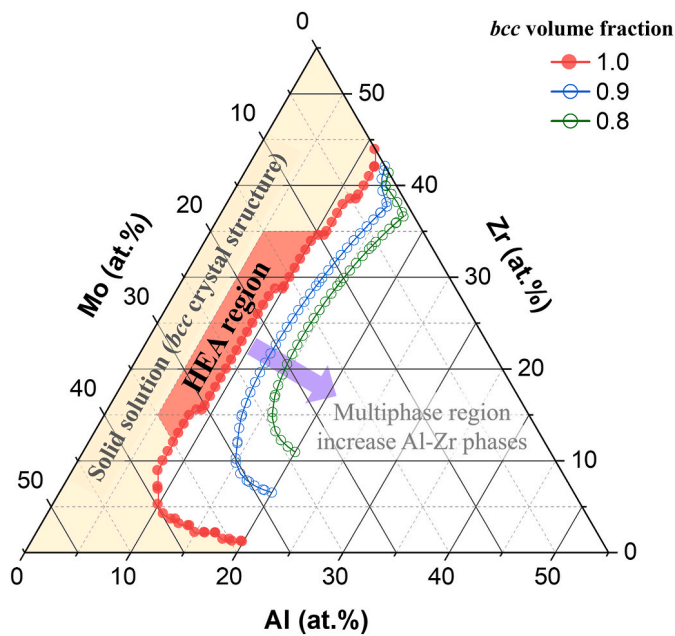

**Fig. 9.** Simulated *bcc*-phase stability regarding the Al, Mo, and Zr components for the *x*Al-*y*Mo-*z*Zr-10Nb35Ti system. Ti- and Nb-content are constant, while the composition variation between Al, Mo and Zr sum up the 55 at.% in the alloy: The lines determine the evolution of *bcc*-phase estimated at 1300 °C inside the diagram.

largely due to the limited number of HEAs mapped within this framework and the requirement of forming a single-phase microstructure. The four selected alloys from the AlMoNbTiZr system exhibited low electronic parameters, positioning them within the established bcc-phase region of conventional Ti alloys in the diagram. However, chemical affinities between constituent elements must also be considered to avoid alloy systems prone to multi-phase formation. In this study, three alloys with 20 at.% Al developed intermetallics, as the strong negative mixing enthalpy of Al with Zr, promoted heterogeneity and complex chemical evolution. Only the 10Al15Mo10Nb35Ti30Zr alloy appeared to form a near single-phase microstructure, confirmed by TEM as a *bcc/b2* structure with finely distributed *b2* nano-segregates, and thus considered a solid solution. This alloy displayed high compressive strength, a low elastic modulus, and negligible ductility.

For phase prediction in CCAs, empirical parameters provide only preliminary screening. While their predictive accuracy is limited, they generally indicate the correct trend for solid-solution stability within a given system. Combining several parameters can improve guidance, especially those accounting for chemical affinity, such as $\Delta H_{mix}$ and $\Delta \chi^A$. By contrast, the Calphad method provides more reliable qualitative predictions of phase presence and evolution with composition. However, it may lack precision in defining the quantitative boundary between single-phase and multi-phase regions, due to incomplete ternary and quaternary descriptions.

Our results show that a *bcc/b2* solid solution in the AlMoNbTiZr system is stabilized at high temperatures when Al, Mo, and Zr contents are reduced. Because of the strong negative $\Delta H_{mix}$ between Al and Zr, their concentrations must be limited to suppress intermetallic formation. Experimentally, a "single-phase" microstructure was obtained in the 10Al15Mo10Nb35Ti30Zr alloy, with a measured Al content of 11 at.%. CALPHAD predictions, however, suggest an upper Al limit of ∼8 at.%, independent of Zr and Mo content, for stabilizing a *bcc/b2* solid solution, which is nearly consistent to the observed "single-phase" alloy.

**Declaration of generative AI and AI-assisted technologies in the writing process**

During the preparation of this work the authors used OpenAI service in order to revise the grammar and clarity of the text. After using this tool, the authors reviewed and edited the content as needed, taking full responsibility for the content of the publication.

**Declaration of competing interest**

The authors declare that they have no known competing financial interests or personal relationships that could have appeared to influence the work reported in this paper.

**Acknowledgements**

This research was financially supported by the Czech Science Foundation within the 21–18652 M project. Plasma-arc melting was performed at the MGML facilities (mgml.eu), supported by the Czech Research Infrastructures program under the LM2023065 project. Financial support by the Operational Programme Johannes Amos Comenius of the MEYS of the Czech Republic, within the frame of the project Ferroic Multifunctionalities (FerrMion) [project No. CZ.02.01.01/00/22_008/0004591], co-funded by the European Union is also gratefully acknowledged. Special thanks are due to Dr. D. King for providing the code used in the phi-parameter calculation; and to Dr. J. Veselý for the help and guidance on the TEM analyses.

**Appendix A. Supplementary data**

Supplementary data to this article can be found online at https://doi.org/10.1016/j.jmrt.2025.09.118.

**Data availability**

The data that support the findings of this study are openly available in the Zenodo repository at https://doi.org/10.5281/zenodo.15480835 under the CC-BY 4.0 license.